\documentclass[twocolumn,showpacs,preprintnumbers,amsmath,amssymb,nofootinbib,
tightenlines]{revtex4}
\usepackage{latexsym,epsf}
\usepackage{graphicx}
\usepackage{bm}
\usepackage{psfig,epsfig}
\begin{document}
\preprint{\tighten{\vbox{\hbox{NCU-HEP-k017}
\hbox{Aug 2004}
}}}
%

\def\ap#1#2#3{           {\it Ann. Phys. (NY) }{\bf #1} (19#2) #3}
\def\arnps#1#2#3{        {\it Ann. Rev. Nucl. Part. Sci. }{\bf #1} (19#2) #3}
\def\cnpp#1#2#3{        {\it Comm. Nucl. Part. Phys. }{\bf #1} (19#2) #3}
\def\apj#1#2#3{          {\it Astrophys. J. }{\bf #1} (19#2) #3}
\def\asr#1#2#3{          {\it Astrophys. Space Rev. }{\bf #1} (19#2) #3}
\def\ass#1#2#3{          {\it Astrophys. Space Sci. }{\bf #1} (19#2) #3}

\def\apjl#1#2#3{         {\it Astrophys. J. Lett. }{\bf #1} (19#2) #3}
\def\ass#1#2#3{          {\it Astrophys. Space Sci. }{\bf #1} (19#2) #3}
\def\jel#1#2#3{         {\it Journal Europhys. Lett. }{\bf #1} (19#2) #3}

\def\ib#1#2#3{           {\it ibid. }{\bf #1} (19#2) #3}
\def\nat#1#2#3{          {\it Nature }{\bf #1} (19#2) #3}
\def\nps#1#2#3{          {\it Nucl. Phys. B (Proc. Suppl.) } {\bf #1} (19#2) #3} 
\def\np#1#2#3{           {\it Nucl. Phys. }{\bf #1} (19#2) #3}
\def\npp#1#2#3{           {\it Nucl. Phys. }{\bf #1} (20#2) #3}
\def\pl#1#2#3{           {\it Phys. Lett. }{\bf #1} (19#2) #3}
\def\pll#1#2#3{           {\it Phys. Lett. }{\bf #1} (20#2) #3}
\def\pr#1#2#3{           {\it Phys. Rev. }{\bf #1} (19#2) #3}
\def\prr#1#2#3{           {\it Phys. Rev. }{\bf #1} (20#2) #3}
\def\prep#1#2#3{         {\it Phys. Rep. }{\bf #1} (19#2) #3}
\def\prl#1#2#3{          {\it Phys. Rev. Lett. }{\bf #1} (19#2) #3}
\def\prll#1#2#3{          {\it Phys. Rev. Lett. }{\bf #1} (20#2) #3}

\def\pw#1#2#3{          {\it Particle World }{\bf #1} (19#2) #3}
\def\ptp#1#2#3{          {\it Prog. Theor. Phys. }{\bf #1} (19#2) #3}
\def\jppnp#1#2#3{         {\it J. Prog. Part. Nucl. Phys. }{\bf #1} (19#2) #3}

\def\rpp#1#2#3{         {\it Rep. on Prog. in Phys. }{\bf #1} (19#2) #3}
\def\ptps#1#2#3{         {\it Prog. Theor. Phys. Suppl. }{\bf #1} (19#2) #3}
\def\rmp#1#2#3{          {\it Rev. Mod. Phys. }{\bf #1} (19#2) #3}
\def\zp#1#2#3{           {\it Zeit. fur Physik }{\bf #1} (19#2) #3}
\def\fp#1#2#3{           {\it Fortschr. Phys. }{\bf #1} (19#2) #3}
\def\Zp#1#2#3{           {\it Z. Physik }{\bf #1} (19#2) #3}
\def\Sci#1#2#3{          {\it Science }{\bf #1} (19#2) #3}

\def\n.c.#1#2#3{         {\it Nuovo Cim. }{\bf #1} (19#2) #3}
\def\r.n.c.#1#2#3{       {\it Riv. del Nuovo Cim. }{\bf #1} (19#2) #3}
\def\sjnp#1#2#3{         {\it Sov. J. Nucl. Phys. }{\bf #1} (19#2) #3}
\def\yf#1#2#3{           {\it Yad. Fiz. }{\bf #1} (19#2) #3}
\def\zetf#1#2#3{         {\it Z. Eksp. Teor. Fiz. }{\bf #1} (19#2) #3}
\def\zetfpr#1#2#3{         {\it Z. Eksp. Teor. Fiz. Pisma. Red. }{\bf #1} (19#2) #3}
\def\jetp#1#2#3{         {\it JETP }{\bf #1} (19#2) #3}
\def\mpl#1#2#3{          {\it Mod. Phys. Lett. }{\bf #1} (19#2) #3}
\def\ufn#1#2#3{          {\it Usp. Fiz. Naut. }{\bf #1} (19#2) #3}
\def\sp#1#2#3{           {\it Sov. Phys.-Usp.}{\bf #1} (19#2) #3}
\def\ppnp#1#2#3{           {\it Prog. Part. Nucl. Phys. }{\bf #1} (19#2) #3}
\def\cnpp#1#2#3{           {\it Comm. Nucl. Part. Phys. }{\bf #1} (19#2) #3}
\def\ijmp#1#2#3{           {\it Int. J. Mod. Phys. }{\bf #1} (19#2) #3}
\def\ic#1#2#3{           {\it Investigaci\'on y Ciencia }{\bf #1} (19#2) #3}
\def\tp{these proceedings}
\def\pc{private communication}
\def\ip{in preparation}
\relax
\newcommand{\TeV}{\,{\rm TeV}}
\newcommand{\GeV}{\,{\rm GeV}}
\newcommand{\MeV}{\,{\rm MeV}}
\newcommand{\keV}{\,{\rm keV}}
\newcommand{\eV}{\,{\rm eV}}
\newcommand{\Tr}{{\rm Tr}\!}
\renewcommand{\arraystretch}{1.2}
\newcommand{\be}{\begin{equation}}
\newcommand{\ee}{\end{equation}}
\newcommand{\bea}{\begin{eqnarray}}
\newcommand{\eea}{\end{eqnarray}}
\newcommand{\ba}{\begin{array}}
\newcommand{\ea}{\end{array}}
\newcommand{\bc}{\begin{center}}
\newcommand{\ec}{\end{center}}
\newcommand{\bmat}{\left(\ba}
\newcommand{\emat}{\ea\right)}
\newcommand{\bds}{\begin{description}}
\newcommand{\eds}{\end{description}}
\newcommand{\refs}[1]{(\ref{#1})}
\newcommand{\ler}{\stackrel{\scriptstyle <}{\scriptstyle\sim}}
\newcommand{\ger}{\stackrel{\scriptstyle >}{\scriptstyle\sim}}
\newcommand{\lag}{\langle}
\newcommand{\rag}{\rangle}
\newcommand{\ns}{\normalsize}
\newcommand{\cm}{{\cal M}}
\newcommand{\gr}{m_{3/2}}
\newcommand{\p}{\partial}
\newcommand{\bsg}{$b\rightarrow s + \g$}
\newcommand{\Bsg}{$B\rightarrow X_s + \g$}
\newcommand{\atal}{{\it et al.}}
\newcommand{\cq}{{\cal Q}}
\newcommand{\cqt}{{\widetilde {\cal Q}}}
\newcommand{\wtlc}{{\widetilde C}}
\def\321{$SU(3)\times SU(2)\times U(1)$}
\def\tl{{\tilde{l}}}
\def\tL{{\tilde{L}}}
\def\bd{{\overline{d}}}
\def\tL{{\tilde{L}}}
\def\a{\alpha}
\def\b{\beta}
\def\bsg{$ b \rightarrow s + \g$}
\def\g{\gamma}
\def\c{\chi}
\def\d{\delta}
\def\D{\Delta}
\def\db{{\overline{\delta}}}
\def\Db{{\overline{\Delta}}}
\def\e{\epsilon}
\def\f{\frac}
\def\tn{-\frac{2}{9}}
\def\tt{\frac{2}{3}}
\def\l{\lambda}
\def\n{\nu}
\def\m{\mu}
\def\nt{{\tilde{\nu}}}
\def\p{\phi}
\def\P{\Phi}
\def\k{\kappa}
\def\x{\xi}
\def\r{\rho}
\def\s{\sigma}
\def\t{\tau}
\def\th{\theta}
\def\ne{\nu_e}
\def\nm{\nu_{\mu}}
\def\snui{\tilde{\nu_i}}
\def\la{{\makebox{\tiny{\bf loop}}}}
\def\ti{\tilde}
\def\ssc{\scriptscriptstyle}
\def\wtl{\widetilde}
\def\mp{\marginpar}
\def\und{\underline}
\renewcommand{\Huge}{\Large}
\renewcommand{\LARGE}{\Large}
\renewcommand{\Large}{\large}

\title{Some Novel Contributions to Radiative B Decay in Supersymmetry without $R$-parity}

\author{\bf  Otto C. W. Kong and Rishikesh D. Vaidya \\
{\ns\it  Department of Physics, National Central University,
Chung-Li, Taiwan   32054}}

\begin{abstract}
We present a systematic analysis at the leading log order of the influence of 
combination of bilinear and trilinear $R$-parity violating couplings on the decay
\bsg\ . Such contributions have never been explored in the context of \bsg\ decay. 
We show that influence of charged-slepton-Higgs mixing mediated loops can 
dominate the SM and MSSM contributions and hence can provide strong 
bounds on the combination of 
bilinear-trilinear $R$-parity violating couplings. Such contributions are also
enhanced by large $\tan \b$. With substantially extended basis
of operators (28 operators), we provide illustrative analytical formulae of the major
contributions to complement our complete numerical results which demonstrate the 
importance of QCD running effects.
\end{abstract}
\maketitle

{\it Introduction.---} 
The minimal supersymmetric standard model (MSSM) has been the most
popular candidate theory for physics beyond the Standard Model (SM) for the
last couple of decades. With the recent accumulation of evidence for neutrino 
oscillations, it is clear that the lepton number conserving MSSM has to be amended.
The simplest option is then to give up imposing $R$-parity and hence admit 
all gauge invariant terms in a (generic) supersymmetric SM. Other alternatives
include incorporating any particular neutrino sector model such as the seesaw
mechanism with extra gauge singlet superfields. We focus on the first 
option\cite{004,nu}. The model has the special merit that the parameters that
give rise to neutrino masses and mixings also have interesting phenomenological
consequence in the quark and charged lepton sectors. Here in this letter,
we report on some of the novel contributions to the \bsg\ decay from the model.

Within the SM, the flavor sector still needs more scrutiny from theory as well as
experiments.
In particular,
flavor changing neutral current (FCNC) processes are widely considered to be the 
window of the physics beyond SM. Among the processes,  \Bsg\ is a particularly 
attractive candidate. 
The most up-to-date SM
prediction \cite{gam-mis-01} gives
\be 
\label{br-th}
Br\!\left[B \rightarrow X_s + \g \;\mbox{\tiny ($E_{\g} > 1.6$ GeV)}
\right]_{\ssc \mathrm{\tiny SM}} \!\!=\!\!
(3.57 \pm 0.30) \times 10^{-4} \,,
\ee
while the experimental number (world-average) is 
\cite{bsg-exp}
\be
\label{br-exp}
Br\!\left[B \rightarrow X_s + \g \;\mbox{\tiny ($E_{\g} > 1.6$ GeV)}
\right]_{\ssc \mathrm{\tiny EXP}} \!\!=\!\! (3.34 \pm 0.38) \times 10^{-4}\,.
\ee
It clearly leaves not much room for new physics contributions. Hence, can be used to
obtain stringent constraints on flavor parameters of various new physics models. 

There have been some studies on the process within the general framework of 
$R$-parity violation \cite{carlos,besmer}. 
Ref.\cite{carlos},  fails to consider the
additional 18 four-quark operators which, in fact, give the dominant contribution
in most of the cases. The more recent work of ref.\cite{besmer} has considered a 
complete operator basis. However, we find their formula for Wilson coefficient incomplete.
In fact, the particular type of contributions --- namely, one from a combination of
a bilinear and a trilinear $R$-parity violating (RPV) parameters, we focus on here,
has not been studied in any detail before.

It is not possible for us to give much analytical details of our study here in this short 
letter. We will only outline the major features of the full analysis given in a parallel 
report \cite{014}, to which interested readers are referred. 
We adopt an optimal phenomenological parametrization of  the full model 
Lagrangian, dubbed the single single-VEV parametrization (SVP), first
explicitly advocated in refs.\cite{ru12}. It is essentially about choosing a basis
for Higgs and lepton superfields in which all the ``sneutrino" VEVs vanish.
The formulation gives the simplest
expressions for all the mass matrices of the fermions and scalars without 
{\it a priori} assumption on the admissible form of $R$-parity violation. In
particular, all the RPV effects to the fermion mass matrices are characterized
by the three bilinear parameter $\mu_i$'s. Working under the formulation, it has
been pointed out in ref.\cite{as5} that there are interesting contributions to the
down squark and charged slepton mass matrices of the form 
$(\, \mu_i^*\lambda^{\!\prime}_{ijk}\, )$  and
$(\, \mu_i^*\lambda_{ijk}\, )$. These give explicit indications of the
existence of bilinear-trilinear type contributions to fermion dipole moments at
1-loop order, including the transitional moment term to be identified with the \bsg\
decay. Detailed analytical and numerical studies have been performed on the
case of neutron electric dipole moment\cite{as4-as6} and $\mu \to e + \g$ \cite{as7}.
Here, we report on the more difficult calculation of \Bsg\ . Background details 
on the model and the various mass matrices are given in ref.\cite{004}. Based on
the latter results, we implement our (1-loop) calculations using mass eigenstate 
expressions\cite{014}, hence free from the commonly adopted mass-insertion 
approximation. While a trilinear RPV parameter gives a coupling, a bilinear 
parameter now contributes only through mass mixing matrix elements characterizing 
the effective couplings of the mass eigenstate running inside the loop. The $\mu_i$'s
are involved in fermion, as well as scalar mixings\cite{004}. There are also the
corresponding soft bilinear  $B_i$ parameters involved only in scalar 
mixings\cite{004}. Combinations of $\mu_i$'s and  $B_i$'s with the trilinear
$\lambda^{\!\prime}_{ijk}$ parameters are our major focus. 

{\it The Effective Hamiltonian Approach.---}
The partonic transition \bsg\ is described by the magnetic penguin diagram.
Under the effective Hamiltonian approach, the corresponding Wilson 
coefficients of the standard $\cq_7$ operator has many RPV contributions at
the scale $M_{\!\ssc W}$. For example, we separate the contributions from
different type of diagrams as 
$C_7 = C^{\ssc W}_7 \,+ \,C^{\ssc \tilde g}_7 \, + \,C^{\ssc \chi^-}_7 \, 
+\, C^{\ssc \chi^{\mbox{\tiny 0}}}_7 \, +\, C^{\ssc \phi^-}_7 \,
+\, C^{\phi^{\mbox{\tiny 0}}}_7$. The neutral scalar loop contribution
$C^{\phi^{\mbox{\tiny 0}}}_7$, as an illustrative case, is proportional to
\footnotesize\[
{{\cal Q}_{\!{d}} \over M_{\!\scriptscriptstyle S_m}^2} \left[
\widetilde{\cal N}_{\!\scriptscriptstyle nmj}^{\!\scriptscriptstyle R}\,
 \widetilde{\cal N}_{\!\scriptscriptstyle nmi}^{\!\scriptscriptstyle L^*} \;
{{m}_{\!\scriptscriptstyle d_n} \over m_{\!\scriptscriptstyle d_j}} \;
F_3\!\!\left({ m_{\!\scriptscriptstyle d_{n}}^2 \over 
M_{\!\scriptscriptstyle S_m}^2} \right) 
+\widetilde{\cal N}_{\!\scriptscriptstyle nmj}^{\!\scriptscriptstyle L}\,
 \widetilde{\cal N}_{\!\scriptscriptstyle nmi}^{\!\scriptscriptstyle L^*} \;
F_1\!\!\left({ m_{\!\scriptscriptstyle d_{n}}^2 \over 
M_{\!\scriptscriptstyle S_m}^2} \right)\right]\;,
\] \normalsize
where the effective vertex couplings 
$ \widetilde{\cal N}^{{\!\scriptscriptstyle L}^*}_{\scriptscriptstyle nmi}$
and
$ \widetilde{\cal N}^{{\!\scriptscriptstyle R}^*}_{\scriptscriptstyle nmi}$
for the mass eigenstates each contains the a $\l'$-coupling contribution\cite{014}.

Apart from the 8 SM operators with additional contributions, 
we actually   have  to consider many more operators with admissible nonzero  Wilson 
coefficients at  $M_{\!\ssc W}$ resulting from the RPV couplings. These are the
chirality-flip counterparts $\cqt_7$ and $\cqt_8$ of the standard (chromo)magnetic 
penguins $\cq_7$ and $\cq_8$, and a whole list of 18 new relevant four-quark 
operator of current-current type to be given as :
\bea
{\cal Q}_{9-11}
\!\! & =& \!\!
\left({\bar s}_{L\a}\,\g^{\m} \, b_{L\b}\right)
\, \left({\bar q}_{R\b}\,\g_{\m}\,q_{R\a}\right)\,,~q = d,s,b;  \\
{\widetilde {\cal Q}_{3,4}}
\!\! & =& \!\!
\left({\bar s}_{R\a}\,\g^{\m}b_{R\a,\b}\right)
\,\sum_{i=u,c,d,s,b}\,\left({\bar q}_{Ri\b}\,\g^{\m}\,q_{Ri\b,\a}\right)\,;  \\
{\widetilde {\cal Q}_{5,6}}
\!\! & =& \!\!
 \left({\bar s}_{R\a}\,\g^{\m}b_{R\a,\b}\right)
\,\sum_{i=u,c,d,s,b}\,\left({\bar q}_{Li\b}\,\g^{\m}\,q_{Li\b,\a}\right)\,;\\
{\widetilde{\cal Q}}_{ 9-13} 
\!\! & =& \!\!
 \left({\bar s}_{R\a}\,\g^{\m}\,b_{R\b}\right)
\, \left({\bar q}_{L\b}\, \g_{\m}\,q_{L\a}\right)\,, ~q = d,s,b,u,c;  
\eea
and six more operators from $\l''$ couplings\cite{014} we skip here for brevity. 
The interplay among the full set of 28 operators is what 
makes the analysis complicated. The effect of the QCD corrections proved to be 
very significant even for the RPV parts.

We skip here the details involved in the evaluation of the various
effective Wilson coefficients for the decay rate of \bsg\ and give only the
numerical results from our leading log (LL) order analysis \cite{014}  :
\begin{eqnarray}
\label{c_mb}
C_7^{\mathrm{eff}}(m_b)&=& 
-0.351 \, C_{2}^{\mathrm{eff}}(M_{\!\ssc W})
+0.665 \, C_{7}^{\mathrm{eff}}(M_{\!\ssc W}) \nonumber\\
&&+0.093 \, C_{8}^{\mathrm{eff}} (M_{\!\ssc W})
-0.198 \, C_{9}^{\mathrm{eff}}(M_{\!\ssc W}) \nonumber \\
&&-0.198  \, C_{10}^{\mathrm{eff}} (M_{\!\ssc W})
-0.178  \, C_{11}^{\mathrm{eff}}(M_{\!\ssc W})\;, \nonumber \\
[0.5cm]
{\wtl C}_7^{\mathrm{eff}}(m_b)&=&
0.381  \, {\wtl C}_{1}^{\mathrm{eff}}(M_{\!\ssc W})
+0.665  \, {\wtl C}_{7}^{\mathrm{eff}}(M_{\!\ssc W})\nonumber\\
&&+0.093  \, \tilde C_{8}^{\mathrm{eff}}(M_{\!\ssc W})
-0.198  \, {\wtl C}_{9}^{\mathrm{eff}}(M_{\!\ssc W}) \nonumber \\
&&-0.198  \, {\wtl C}_{10}^{\mathrm{eff}}(M_{\!\ssc W})
-0.178  \, {\wtl C}_{11}^{\mathrm{eff}}(M_{\!\ssc W})\nonumber\\
&&+0.510  \, {\wtl C}_{12}^{\mathrm{eff}}(M_{\!\ssc W})
+0.510 \, {\wtl C}_{13}^{\mathrm{eff}}(M_{\!\ssc W})\nonumber\\
&&+0.381 \, {\wtl C}_{14}^{\mathrm{eff}}(M_{\!\ssc W})
-0.213  \, {\wtl C}_{16}^{\mathrm{eff}}(M_{\!\ssc W})\;.
\end{eqnarray}

The branching fraction for $Br (b \rightarrow s + \g )$  is expressed through the 
semi-leptonic decay $b \rightarrow u|c e{\bar \nu}$ so that the large
bottom mass dependence $( \sim m^5_b)$ and uncertainties in CKM elements
cancel out.
\be
Br (b \rightarrow s + \g) = \frac{\Gamma (b \rightarrow s + \gamma )}
{\Gamma (b \rightarrow u|c \,e\,{\bar \nu_e})} \;
Br_{\mathrm{exp}} (b \rightarrow u|c \,e\,{\bar \nu_e})\;,
\ee
where $Br_{\mathrm{exp}} (b \rightarrow u|c\,e\,{\bar \nu_e}) = 10.5 \%$
and
\be
\Gamma (b \rightarrow s \g)  =  \f{\a \, m^5_b}{64 \pi^4} \,
\left(\,|C^{\mathrm{eff}}_7 (\m_b)|^2 + 
|\widetilde{C}^{\mathrm{eff}}_7(\m_b)|^2 \right) \;.
\ee
Note that we have also to include RPV contributions to the semi-leptonic rate
for consistency\cite{014}.

{\it Analytical  Appraisal of the Results. ---} 
There are three kinds of bilinear RPV parameters, $\m_i,B_i$ and 
${\tilde m}^2_{{\ssc L}_{0i}}$ related by the tadpole equation constraints\cite{as5,004}.
Without loss of generality, we choose $\m_i$ and $B_i$ to be independent. 
The influence of a $|B_i|$ (or $|\m_i|$), in conjunction with $|\l^{'}_{ijk}|$ is felt
through the lepton number violating mass mixings in (s)leptonic propagators of tree and 
penguin diagrams. $|B_i|$ insertions may have much stronger influence than the 
$|\m_i|$ as the former case is inversely proportional to light slepton mass-squared 
whereas the latter ones come with the inverse of heavier squark mass-squared. We focus our 
discussion here on the $|B_i|$ insertions to provide an analytical appraisal of the numerical 
results. The case for $\m_i$  can be appreciated in a similar fashion\cite{nedm}.

There are two kinds of $B_i$-$\l'$ combinations that contribute to \bsg\ at 1-loop: 
(a) $B^*_i \l^{'}_{ij2}$, and (b) $B_i \l^{'*}_{ij3}$. These involve quark-scalar loop diagrams.
Case (a) leads to the $b_{\!\ssc L} \rightarrow s_{\!\ssc R}$ transition (where SM and MSSM 
contribution is extremely suppressed) whereas case (b) leads to SM-like $b_{\!\ssc R} 
\rightarrow s_{\!\ssc L}$ transition. For the purpose of illustration, we will assume a 
degenerate slepton spectrum and take the sleptonic index $i=3$ as a representative. 
The $j=3$ [$2$] contributions for case (a) [(b)] with both sneutrino-Higgs mixings
and charged-slepton-Higgs mixings are easy to appreciate. For the $j$ values,
the charged loop contributions are still possible by invoking CKM mixings. 
Consider the contribution of case (a) with $|B_3^* \l'_{332}|$ to the Wilson coefficient 
${\wtl C}_7$, for instance. Through the extraction of the bilinear mass mixing
effect under a perturbative diagonalization of the mass matrices\cite{004}, we obtain
 \begin{widetext}
\be
{\wtl C}^{\ssc \phi^-}_7   \approx 
 \frac{ - |V^{tb}_{\!\mbox{\tiny CKM}}|^2 \,|B_3^* \l'_{332}| }
{M^2_s}
\left\{
y_b  \, \tan \b
 \left[ 	F_2 \!\!  	\left(
{{m}_{\!\scriptscriptstyle t}^2 \over M_{\!\scriptscriptstyle \tilde{\ell}}^2}
\right) 
+ {\cal Q}_u \,		F_1 \!\! 	\left(
{{m}_{\!\scriptscriptstyle t}^2 \over M_{\!\scriptscriptstyle \tilde{\ell}}^2}
\right)		\right] 
+ \! \f{y_t \, m_t}{m_b} 
\left[ 	F_4 \!\! 	\left(
{{m}_{\!\scriptscriptstyle t}^2 \over M_{\!\scriptscriptstyle \tilde{\ell}}^2}
\right) + 
{\cal Q}_u \,	F_3 \!\! \left(
{{m}_{\!\scriptscriptstyle t}^2 \over M_{\!\scriptscriptstyle \tilde{\ell}}^2}
\right)   	\right] 		\right\}\\
\ee
\end{widetext}
\be
{\wtl C}^{\phi^{\mbox{\tiny 0}}}_7  \approx 
\f{-{2\cal Q}_d \, y_b \, |B_3^*  \l'_{332}| \tan \b }{M^2_s M_{\!\scriptscriptstyle S}^2}
F_1 \!\! \left(
{ m_{\!\scriptscriptstyle b}^2 \over 
M_{\!\scriptscriptstyle S}^2}
\right)
\ee
for the  charged and neutral scalar loop, respectively. 
In the above equations, proportionality to $\tan\!{\b}$ shows the 
importance of these contributions in the large $\tan\!{\b}$ limit. The $M_s^2$,
$M_{\!\scriptscriptstyle \tilde{\ell}}^2$, $M_{\!\scriptscriptstyle S}^2$,
are all scalar (slepton/Higgs) mass parameters.
The term proportional to $y_t$ above has chirality flip into the loop. Thinking in terms of
the electroweak states, it is easy to appreciate that the loop diagram giving a corresponding
term for ${\wtl C}^{\phi^{\mbox{\tiny 0}}}_7$ ({\it cf.} involving
$\widetilde{\cal N}_{\!\scriptscriptstyle nm3}^{\!\scriptscriptstyle L}\,
 \widetilde{\cal N}_{\!\scriptscriptstyle nm2}^{\!\scriptscriptstyle R^*}$)
requires a Majorana-like scalar mass insertion, which has to arrive from other RPV 
couplings\cite{004}. In the limit of perfect mass degeneracy between the scalar and 
pseudoscalar part (with no mixing) of multiplet, it vanishes. Dropping this smaller
contribution, together  with the difference among the Inami-Lim loop functions and the
fact that the charged loop has more places to attach the photon (with also larger charge
values) adding up, we expect the ${\wtl C}^{\ssc \phi^-}_7$ to be larger than
${\wtl C}^{\phi^{\mbox{\tiny 0}}}_7$.

\begin{figure}[b]
\includegraphics[scale=0.7]{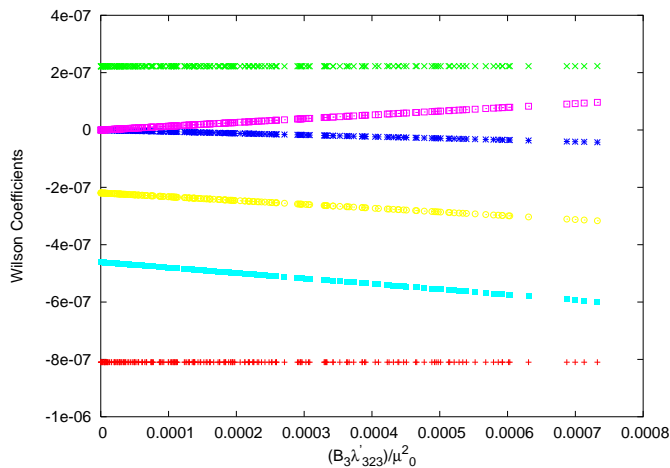}
\caption{\label{btwn-3-323} 
Various Wilson coefficients versus $|B_{\ssc 3}\l'_{\ssc 323}|$.
`+' sign stands for MSSM chargino contribution, `$\times$' stands for the
MSSM charged Higgs contribution, `$\star$' stands for  sneutrino contribution, all
contributing to $C_7$.
`Empty square' stands for $C_{11} (M_W)$, `filled square for $C_7 (M_W)$ and
`empty circle' for $C_7 (m_b)$.}
\end{figure}
\begin{figure}[b]
\includegraphics[scale=0.6]{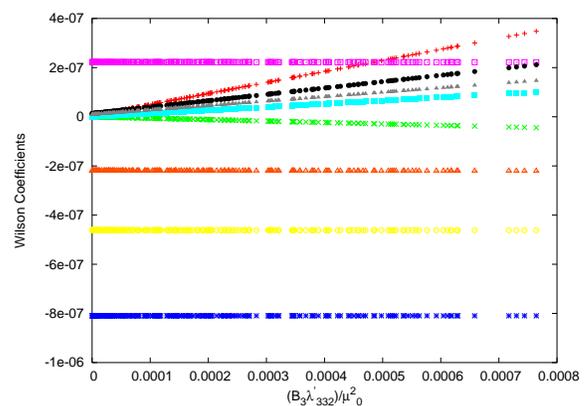}
\caption{\label{btwn-3-332} 
Various Wilson coefficients versus $|B_{\ssc 3}\l'_{\ssc 332}|$.
`+' sign stands for  charged slepton contribution, '$\times$' stands for sneutrino
contribution, `$\star$' stands for MSSM chargino contribution, all contributing to
$\wtl C_7$. `Empty square'
stands for MSSM charged Higgs contribution, `filled square' for $\tilde C_{11} (M_W)$,
`empty circle' for $C_7 (M_w)$, `filled circle' for $\tilde C_7 (M_W)$, `empty triangle'
for $C_7 (m_b)$, and `filled triangle' for $\tilde C_7 (m_b)$.}
\end{figure}

{\it Numerical Results. ---} 
The basic strategy for the implementation of our numerical study is to choose model
parameters in such a way that in the limit that  the RPV parameters vanish, 
the resulting MSSM gives the decay rate  well within the experimental limit\cite{mssm}. 
The otherwise arbitrariness in the choice mostly does not have much of an effect on 
the qualitative dependence of the results on a specific combination of RPV parameters.
Note that we always keep $R$-parity conserving flavor violating squark and slepton 
mixings vanishing, to focus on the RPV effects. We take non-vanishing values
for relevant combinations of a bilinear and a trilinear RPV parameters one at a time,
and stick to real values only. Our model choice is (with all mass dimensions
given in GeV):  squark masses 300, down-type Higgs mass 300, $\m_{\ssc 0} = -300 $ 
sleptons mass 150 and gaugino mass $M_2 = 200$ (with $M_1 = 0.5 M_2$ and 
$M_3 = 3.5M_2$), $\tan\!\b = 37$ and $A$ parameter 300. 
The mass for $H_u$ and soft bilinear parameter $B_0$ are determined
by electroweak symmetry breaking conditions which are modified in the presence of 
RPV parameters \cite{as5,004}.

Under the scenario discussed, we impose the experimental number to obtain bounds
for each combination of RPV parameters independently (given in Table I). 
We 
address here a couple of cases in a bit more detail. Consider, for instance, the case (b)
combination $|B_{\ssc 3}\l^{'*}_{\ssc 323}|$.  We obtain a bound of $5.0 \times 10^{-5}$, when
normalized by a factor of $\m^2_{\ssc 0}$. Since this is a $b_{\!\ssc R} \rightarrow s_{\!\ssc L}$
transition, the RPV contribution interferes with the SM as well as the MSSM contribution.
In Fig.~\ref{btwn-3-323} we have plotted the relevant Wilson coefficients. Over and above
the loop contributions we see that there are contributions coming from four-quark 
operator with Wilson coefficients $C_{11}$ ($\propto y_b$) which is stronger than 
the other two four-quark quark coefficients ${\wtl C}_{10,13} \propto y_s$
(not shown in the graph). Since the neutral scalar loop contribution is proportional
to the loop function $F_1$ (which is or order .01), it is suppressed compared to 
current current contributions. Also here the charged scalar contribution comes only 
with chirality flip inside the loop and has a CKM suppression. So the current-current is 
dominant. It has a more subtle role to play when one writes the regularization
scheme-independent $C^{\mathrm{eff}}_7 = C_7 -C_{11}$ at scale $M_{\ssc W}$. 
Due to dominant and negative
sign chargino contribution ($A_t\, \m_{\ssc 0} < 0$), the positive sign $C_{11}$ interferes
constructively with $C_7$ and enhances the rate.

The case (b) combination $|B_3 \l'_{332}|$ is a different story, as it leads to 
$b_{\!\ssc L} \rightarrow s_{\!\ssc R}$ transition and hence RPV does not interfere 
with SM or MSSM contribution. This leads to a bound of $7.3\times 10^{-3}$ after 
normalization by $\m^2_{\ssc 0}$. In Fig.~\ref{btwn-3-332} we have again plotted 
the relevant Wilson coefficients. Unlike the previous case, here we see that the contributions 
from charged scalar loop dominates over both neutral scalar as well as current-current
contributions. This is in accordance with our analytical expectations. Again, the
current-current contribution due to ${\wtl C}_{11}$ has a very subtle  role to play 
here. The regularization scheme-independent effective Wilson coefficient 
${\wtl C}^{\mathrm{eff}}_7 = {\wtl C}_7 -{\wtl C}_{11}$ at the scale $M_{\ssc W}$. 
The  negative sign leads to cancellations and hence weakens the bound. 

The influence of $|\m_i \l'_{ijk}|$ is less stronger than the $B_i$ insertions. The
$\m_i$ insertions affect the MSSM chargino and the neutralino type of diagrams by
mixing them with the charged and the neutral leptons. Since such contributions are
suppressed by the heavier squark masses, the influence is not as strong as slepton
loops. However there exist gluino mediated loop diagrams with a flavour violating
chirality flip in the down-squark propagator ($\propto |\m_i \l'_{ijk}|$) which gives
non-negligible contributions and indeed lead to good bounds. 
%

\begin{table}[t]
\caption{\bf Bounds for the products of bilinear and trilinear RPV couplings.  }
\label{bounds}
\begin{tabular}{|l|l|l|}\hline
{\bf Product } & {\bf Our bound} & {\bf Wilson Coef.}\\\hline\hline  
$\left|\f{B_i \cdot \l'_{i23}}{\m^2_{\ssc 0}}\right|$& $5.0\times 10^{-5}$&
$C_{7,8},{\wtl C}_{7,8},C_{11},\wtlc_{10},\wtlc_{13}$\\\hline
$\left|\f{B_i \cdot \l'_{i32}}{\m^2_{\ssc 0}}\right|$& $7.4\times 10^{-3}$&
$C_{7,8},{\wtl C}_{7,8},C_{10},\wtlc_{11}$\\\hline
$\left|\f{B_i \cdot \l'^*_{i33}}{\m^2_{\ssc 0}}\right|$& $2.3\times 10^{-3}$&
$C_{7,8},{\wtl C}_{7,8}$\\\hline
$\left|\f{B_i \cdot \l'_{i22}}{\m^2_{\ssc 0}}\right|$& $6.5\times 10^{-2}$&
$\wtlc_{7,8}$\\\hline
$\left|\f{B_i \cdot \l'^*_{i13}}{\m^2_{\ssc 0}}\right|$& $8.0\times 10^{-2}$&
$ C_{7,8}$\\\hline
$\left|\f{B_i \cdot \l'^*_{i12}}{\m^2_{\ssc 0}}\right|$& $4.5\times 10^{-2}$&
$ \wtlc_{7,8}$\\\hline
$\left|\f{\m_i \cdot \l'^*_{i23}}{\m_{\ssc 0}}\right|$& $2.2\times 10^{-3}$&
$C_{7,8}$\\\hline
$\left|\f{\m_i \cdot \l'^*_{i32}}{\m_{\ssc 0}}\right|$& $1.0\times 10^{-2}$&
${\wtl C}_{7,8}$\\\hline
$\left|\f{\m_i \cdot \l'^*_{i33}}{\m_{\ssc 0}}\right|$& $8.0\times 10^{-2}$&
$C_{7,8}$\\\hline
\end{tabular}
\end{table}

{\it Conclusions. ---} To conclude we have systematically studied the influence of the 
combination of bilinear-trilinear RPV parameters on the decay \bsg\ analytically as 
well as numerically. Such a study has not been attempted before.
We demonstrate the plausible dominance of the RPV contributions over conventional SM and 
MSSM contributions in some parameter space regions. These contributions are enhanced 
by large $\tan\!\b$. It is shown that charged-slepton-Higgs mixing mediated loop typically 
dominates over the sneutrino-Higgs mixing mediated loop.  Our study has consistently 
incorporated all the QCD corrections at the leading log order, with a whole list of extra
operators and their Wilson coefficients arising from RPV couplings.
We have shown that, through the formulation of scheme-independent effective Wilson
coefficients, the new current-current operators can considerably influence the decay rate. 
Under a typical and compatible model parameter choice, we obtain strong bounds on several combinations of RPV parameters. Bounds on such bilinear-trilinear parameter combinations
are not available before.  Various \bsg\ contributions over different parameter space regions
may complicate the story and partial cancellation among them are a likely possibility. And
our leading log calculation bears relatively large certainty. Nevertheless, the bounds show
values of the RPV parameter combinations that will play a major role in endangering the
compatibility of the theoretical \bsg\ result with the experimental limits. This interpretation
of our results is quite robust.

Our analytical formulae include all RPV contributions at 1-loop level. Numerical study
has also been performed on combinations of trilinear parameters\cite{014}. We quote here a few
exciting bounds under a similar sparticle spectrum. 
For instance $|\l'_{i33}\cdot\l^{'*}_{i23}|$ for $i=2,3$
should be less than $1.6\times 10^{-3}$ to be compared with rescaled existing bound of
$2\times 10^{-2}$.

{\it Acknowledgment. ---}
The work of O.K. is partially supported by grant number NSC 92-2112-M-008-044 of the 
National Science Council of  Taiwan; R.V. is supported by the same agent under
post-doc grant number NSC 92-2811-M-008-012.


\begin{thebibliography}{99}
\bibitem{004}
For a pedagogical review of the formulation please see
O.C.W.~Kong,  {\it Int.J.Mod.Phy.} A{\bf 19}, 1863  (2004).
\bibitem{nu}
For discussion of neutrino masses under the same formulation and notation here, see
S.K.~Kang and O.C.W.~Kong, {\it Phys. Rev.} {\bf D69},  {\it 013004} (2004);
also  O.C.W.~Kong, {\it Mod. Phy. Lett.} A {\bf 14}, 903 (1999);
K.~Cheung and O.C.W.~Kong,  
Phys. Rev. {\bf D61},  {\it 113012} (2000).
\bibitem{gam-mis-01} 
A. Buras \atal\, \npp{B631}{02}{219};
P. Gambino and M. Misiak, \npp{B611}{01}{338}. 
Also see A. Kagan and M. Neubert, Eur. Phys. J.C {\bf 7}, 5 (1999).
\bibitem{bsg-exp} 
C. Jessop, SLAC-PUB-9610 (2002). See also
S.~Chen \atal\ (CLEO Collaboration), {\it Phys. Rev. Lett.} {\bf 87}, 251807 (2001);
 K.~Abe \atal\ (BELLE Collaboration), \pll{B511}{01}{151}; 
 R.~Barate \atal\ (ALEPH Collaboration) \pl{B429}{98}{169}.
\bibitem{carlos} 
B. de Carlos \atal\ {\it Phys. Rev.} {\bf D55}, {\it 4222} (1997).
\bibitem{besmer} 
Th. Besmer \atal\, {\it Phys. Rev.} {\bf D63} {\it 055007} (2001).
\bibitem{014} 
O.C.W.Kong \atal\ NCU-HEP-k014, hep-ph/0403148.
\bibitem{ru12}
M. Bisset \atal\,
{\it Phys. Lett.} {\bf B430}, 274 (1998);
Phys. Rev. {\bf D62},  {\it 035001} (2000).
\bibitem{as5}
O.C.W. Kong, JHEP {\bf 0009}, {\it 037} (2000).
\bibitem{as4-as6}
Y.-Y. Keum \atal\ ,  {\it Phys.Rev. Lett.} {\bf 86}, 393 (2001);
Phys. Rev. {\bf D63}, {\it 113012} (2001).
\bibitem{as7}
K. Cheung \atal\ , {\it Phys. Rev.} {\bf D64},  {\it 095007} (2001).
\bibitem{mssm}
For MSSM results, see the recent review by T.Hurth, hep-ph/0212304, and references therein.
\bibitem{nedm}
See also a parallel discussion for the $\m_i$ insertions  fermion mixings
in case of quark dipole moment given in Ref.\cite{as4-as6}.

\end{thebibliography}
\end{document}